\documentclass[pdflatex,sn-basic]{sn-jnl}

\usepackage{multirow}
\usepackage{multicol}
\usepackage{subfig}
\usepackage{nicematrix}
\usepackage{makecell}

\DeclareMathOperator*{\argmax}{argmax}

\jyear{2021}%

\theoremstyle{thmstyleone}%
\newtheorem{theorem}{Theorem}
\newtheorem{lemma}{Lemma}
\theoremstyle{thmstyletwo}%

\theoremstyle{thmstylethree}%

\begin{document}

\title[GRAPE-S]{GRAPE-S: Near Real-Time Coalition Formation for Multiple Service Collectives}


\author[1]{\fnm{Grace} \sur{Diehl}}\email{diehlg@oregonstate.edu}

\author*[1]{\fnm{Julie A.} \sur{Adams}}\email{julie.a.adams@oregonstate.edu}

\affil[1]{\orgdiv{Collaborative Robots and Intelligent Systems Institute}, \orgname{Oregon State University}, \orgaddress{
\city{Corvallis}, \postcode{97331}, \state{Oregon}, \country{United States}}}


\abstract{

Robotic collectives for military and disaster response applications require coalition formation algorithms to partition robots into appropriate task teams. Collectives' missions will often incorporate tasks that require multiple high-level robot behaviors or \textit{services}, which coalition formation must accommodate. The highly dynamic and unstructured application domains also necessitate that coalition formation algorithms produce near optimal solutions (i.e., $>95$\% utility) in near real-time (i.e., $<5$ minutes) with very large collectives (i.e., hundreds of robots).  No previous coalition formation algorithm satisfies these requirements. An initial evaluation found that traditional auction-based algorithms' runtimes are too long, even though the centralized simulator incorporated ideal conditions unlikely to occur in real-world deployments (i.e., synchronization across robots and perfect, instantaneous communication). The hedonic game-based GRAPE algorithm can produce solutions in near real-time, but cannot be applied to multiple service collectives. This manuscript integrates GRAPE and a services model, producing GRAPE-S and Pair-GRAPE-S. These algorithms and two auction baselines were evaluated using a centralized simulator with up to 1000 robots, and via the largest distributed coalition formation simulated evaluation to date, with up to 500 robots. The evaluations demonstrate that auctions transfer poorly to distributed collectives, resulting in excessive runtimes and low utility solutions. GRAPE-S satisfies the target domains' coalition formation requirements, producing near optimal solutions in near real-time, and Pair-GRAPE-S more than satisfies the domain requirements, producing \textit{optimal} solutions in near real-time. GRAPE-S and Pair-GRAPE-S are the first algorithms demonstrated to support near real-time coalition formation for very large, distributed collectives with multiple services.

}




\maketitle

\section{Introduction}

Robotic collectives' ability to  perform numerous tasks distributed across large spatial areas efficiently can benefit military and disaster response applications (i.e., surveillance, damage inspections) \cite{DefenseAdvancedResearchProjectsAgency2019OFFensiveTactics,Hildmann2019Review:Safety}. Collectives' missions will often require robot teams that combine multiple high-level behaviors or \textit{services} \cite{Service2011CoalitionAlgorithms}. Consider an urban raid where robots simultaneously maintain overwatch, secure buildings, and search for outdoor hazards \cite{DefenseAdvancedResearchProjectsAgency2019OFFensiveTactics}. Overwatch requires aerial robots with long battery lives (e.g., fixed wings) and long-range sensors, while securing buildings requires agile robots (e.g., quadrotors) with sensors for navigating dense obstacles. A single service cannot represent these tasks, as their required robot capabilities are different. Meanwhile, outdoor search can use either robot type and benefit from the cooperation of both. Leveraging collectives for such missions requires effective \textit{coalition formation for task allocation}, assigning robots to appropriate task teams and determining which service each robot will perform. Coalition formation must be achieved in real-time (i.e., seconds) or near real-time (i.e., $<5$ minutes) to enable collectives to adapt to the rapid task and environmental changes typical of the application domains.

Real-time or near real-time collective coalition formation is challenging, due to computational complexity \cite{Sandholm1999} and collectives' scale ($>50$ robots, potentially thousands) \cite{Brambilla2013,Diehl2022ScalableCollectives,Hamann2018SwarmApproach}. Additionally, the application domains' communication constraints must be considered. Military and disaster response operations often lack permanent communication infrastructure (e.g., cellular) \cite{Jahir2019RoutingSurvey,Legendre201130Research,Shah2019TowardsNetworks}. Temporary, deployed networks (e.g., ad hoc networks) can be used instead or to supplement, but have less power and bandwidth, making them easily overwhelmed by operational demands \cite{Muralidhar2018AnNetworks}. Algorithms must minimize communication in order to reserve this limited bandwidth for mission-critical messages. Another property of deployed networks is that network nodes have relatively limited range \cite{Muralidhar2018AnNetworks}. Messages propagated across large distances must pass through many nodes, introducing delays and increasing the likelihood of lost messages; thus,  frequent communication with a centralized entity is infeasible, and distributed, local communication is preferred.

Prior robot coalition formation research is extensive but has focused primarily on smaller multiple robot systems (e.g., \cite{Service2011CoalitionAlgorithms,Sujit2008UAVFormation,Tang2005ASyMTRe:Reconfiguration}), not collectives. Additionally, early attempts at scalable coalition formation generally required communication with a central entity (e.g., \cite{Mouradian2017ADisasters,Sen2013SA-ANT:Formation,Yeh2016SolvingOptimization}), which is not well-suited to the domain constraints. Recently, auction and hedonic game-based coalition formation algorithms have been identified as potentially viable for domain-constrained collective coalition formation \cite{Diehl2022ScalableCollectives,Jang2018AnonymousSystem}.

The decentralized simultaneous descending auction (SDA) algorithm produces empirically near-optimal team allocations using relatively little communication \cite{Diehl2022ScalableCollectives,Service2014AAllocation}; however, its runtimes can be longer than near real-time (i.e., $>$ 25 min) when individual robots offer multiple services \cite{Diehl2022ScalableCollectives}. Similarly, the auction Robot Allocation through Coalitions Using Heterogeneous Non-Cooperative Agents with a Dynamic Threshold (RACHNA$_{dt}$) algorithm is decentralized and produces empirically optimal team allocations, but has excessive runtimes with multiple service robots (i.e., $>2$ hours) \cite{Diehl2022ScalableCollectives,Vig2006Market-basedFormation}. Other auctions have not been evaluated with large collectives (i.e., $>100$ robots) \cite{Diehl2022ScalableCollectives}.

The hedonic game GRoup Agent Partitioning and Placing Event (GRAPE) algorithm also partially satisfies domain constraints. GRAPE is distributed and produces optimal allocations for 1000 robots in $<5$ min \cite{Diehl2022ScalableCollectives,Jang2018AnonymousSystem}; however, GRAPE applies only when all robots provide a single service \cite{Vig2007CoalitionRobots}. Addressing the coalition formation needs of the military and disaster response domains requires auctions' ability to handle multiple services and GRAPE's runtimes.

This manuscript introduces the GRAPE-S and Pair-GRAPE-S algorithms, which integrate GRAPE with a services model, thus leveraging GRAPE's speed for multiple service coalition formation. Centralized simulation-based evaluations with up to 1000 robots and distributed simulation-based evaluations with up to 500 robots demonstrate that GRAPE-S with sufficient communication resources produces optimal solutions in near real-time for most multiple service collectives. Pair-GRAPE-S performs even better, producing optimal solutions for all of the coalition formation problems considered. The novel algorithms' performance is comparable in solution quality to two common auction protocols, but GRAPE-S and Pair-GRAPE-S have substantially faster worst-case runtimes, achieving near real-time coalition formation. These algorithms are the first demonstrated to support near real-time coalition formation for distributed multiple service collectives at the scale considered.

\section{Problem Formulation}

\textit{Coalition formation for task allocation} partitions agents (e.g., robots) into task teams, called \textit{coalitions} \cite{Service2011CoalitionAlgorithms}. A problem instance comprises a set of $n$ agents, $A = \{a_1, \dots, a_n\}$, and $m$ tasks, $T = \{t_1, \dots, t_m\}$. Task $t_j$ has a utility $u_j$, which is provided by the system user and represents $t_j$'s inherent value to the overall mission \cite{Service2011CoalitionAlgorithms,Zhang2013ConsideringAllocation}. The collective is awarded $u_j$ only if $t_j$ can be completed by its assigned coalition. The optimal solution is the maximum utility set of coalition/task assignments, or \textit{coalition structure} (CS):

\begin{equation}
\begin{gathered}
  CS^* = \underset{CS}{\operatorname{argmax}} \sum_{j=1}^m v_j,\\
\text{where  } v_j = \begin{cases}
u_j &\text{$t_j$ can be completed by its coalition}\\
0 &\text{otherwise.}
\end{cases}
\end{gathered}
\end{equation}

Coalitions may not overlap (i.e., a robot cannot join multiple coalitions), as robot tasks are typically tied to physical locations, and robots can be in only one location at a time. Additionally, tasks are independent of one another and assigned coalitions immediately without long-term planning, due to the highly dynamic and uncertain application domains. Coalition formation with these stipulations is also known as Single-Task Robots, Single-Robot Tasks, Instantaneous Assignment (ST-MR-IA) task allocation. \cite{Gerkey2004ASystems}.

ST-MR-IA task allocation is NP-hard \cite{Sandholm1999,Service2011CoalitionAlgorithms}. Additionally, no polynomial time algorithm can produce a factor $O(\lvert C\rvert^{1-\epsilon})$ or $O(m^{1-\epsilon})$ approximation, where $C$ is the set of non-zero valued coalitions and $\epsilon > 0$, unless P$=$NP \cite{Sandholm1999,Service2011CoalitionAlgorithms}. 

\subsection{The Services Model}

A fundamental challenge of ST-MR-IA coalition formation is efficiently determining whether a coalition is capable of completing its assigned task. Several different models can represent a coalition's capabilities and incorporate constraints, such as robots' hardware capabilities not being instantaneously transferable (e.g., \cite{Tang2005ASyMTRe:Reconfiguration,Vig2007CoalitionRobots,Service2011CoalitionAlgorithms}). This manuscript employs the \textit{services model}, a common coalition formation abstraction where high-level behaviors are considered, and each robot performs a single service at a time \cite{Vig2007CoalitionRobots}.

Specifically, each robot $a_i$ has a service vector $S_{a_i}=(s_1, \dots, s_p)$, denoting the services it can provide. Each task $t_j$ also has a service vector $S_{t_j}=(s_1, \dots, s_p)$, representing the number of each service type it requires. The collective is awarded utility $u_j$ only if $t_j$'s assigned coalition possesses sufficient services, where a robot can perform only one service for one task \cite{Vig2007CoalitionRobots}.

The advantage of the services model is that it implicitly incorporates sensor/actuator configurations, making it efficient when multiple sensors and actuators can produce the same behavior. The abstraction also decreases the level of robotics expertise required to specify tasks, which can make the collective easier for non-developers (e.g., first responders) to command.

\section{Related Work}

Real-time or near real-time coalition formation for collectives that need to be deployed in the real world is relatively unstudied. Although many coalition formation algorithms for software multi-agent and smaller multiple robot systems (i.e., 2-50 robots) achieve 
high-quality task allocations in near real-time (e.g., \cite{Mouradian2017ADisasters,Oh2017Market-BasedEnvironments,Shehory1995TaskAgents}), these algorithms do not transfer directly to collectives (i.e., $>50$ robots). Software multi-agent coalition formation does not account for embodied robots' physical constraints \cite{Vig2007CoalitionRobots}, while multiple robot coalition formation algorithms can be unsuited to larger systems. Multiple robot algorithms that limit coalition size will not scale to collectives, as collective coalitions can have hundreds of robots \cite{Aziz2021Multi-robotApproximation,Vig2006Multi-robotFormation,Zhang2013ConsideringAllocation}. Additionally, centralized multiple robot coalition formation algorithms are incompatible with deployed networks \cite{Agarwal2014Non-additiveFormation,Chen2011ResourceMethodology,Mouradian2017ADisasters,Sen2013SA-ANT:Formation,Yeh2016SolvingOptimization}. The most transferable coalition formation algorithms have been previously identified as multiple robot auctions and hedonic games \cite{Diehl2022ScalableCollectives}.

\subsection{Auction-Based Coalition Formation}

Auctions are a popular coalition formation approach, in which buyers and sellers exchange information about the price of goods \cite{Phelps2008AuctionsLearning}. Algorithms vary by auction protocol and the mapping of tasks and robots to buyers and sellers. 

Many auction protocols and multiple robot system mappings do not translate well to collectives. Robots in first-price, one round auctions bid directly on tasks one at a time in order of receipt (e.g., \cite{Chen2011ResourceMethodology,Gerkey2002,Sujit2008UAVFormation}). These auctions can be fast, but perform arbitrarily poorly, depending on the task ordering. Double auctions (e.g., \cite{Guerrero2017Multi-robotSolutions,Xie2018AFormation}) address this limitation by having robots bid on all tasks at once. Another approach has robots bid through elected project managers that select coalitions from their network neighbors  \cite{Oh2017Market-BasedEnvironments}, which severely restricts the coalitions that can form. Finally, the Automated Synthesis of Multi-robot Task solutions through software Reconfiguration (ASyMTRe) algorithms is incompatible with the services model \cite{Tang2007AAllocation,Zhang2012IQ-ASyMTRe:Tasks}.

The most relevant auction algorithms incorporate \textit{combinatorial auctions}, in which buyers bid on bundles of goods \cite{Vig2006Market-basedFormation}. Two combinatorial auctions, Robot Allocation through Coalitions Using Heterogeneous Non-Cooperative Agents (RACHNA) and simultaneous descending auction (SDA), partially satisfy near real-time domains' requirements in a centralized simulator \cite{Diehl2022ScalableCollectives}.

\subsubsection{Robot Allocation through Coalitions Using Heterogeneous Non-Cooperative Agents (RACHNA)} RACHNA incorporates an \textit{ascending auction} protocol, where bids increase as the auction progresses \cite{Vig2006Market-basedFormation}. The sellers are mapped to robots representing each service type, or \textit{service agents}. A service agent tracks the wages (i.e., rewards for coalition membership) of all robots with its service. The buyers are \textit{task agents} that bid on service bundles. A task agent is awarded a bundle if the bid is at least the robots' current wages, plus $\lvert S_{t_j}\rvert\times \epsilon_i$, where $\lvert S_{t_j}\rvert$ is task $t_j$'s required services and $\epsilon_i$ is a fixed wage increase. A tasks' maximum bid is its utility. The computational complexity is $O(mn^2\log n (u_{max}/\epsilon_i))$, where $u_{max}$ is the largest task utility \cite{Diehl2022ScalableCollectives}. The communication complexity is $O(n^2m\lvert S\rvert(u_{max}/\epsilon_i))$, where $\lvert S\rvert $ is the number of services \cite{Diehl2022ScalableCollectives}.

RACHNA is decentralized and permits multiple services, partially fulfilling collective's coalition formation requirements. However, RACHNA cannot assign coalitions to tasks with $\lvert S_{t_j}\rvert \times\epsilon_i > u_j$, even when sufficient robots exist \cite{Vig2007CoalitionRobots,Vig2006Market-basedFormation,Sen2013AFormation}, which is incompatible with collectives requiring large coalitions \cite{Diehl2022ScalableCollectives}.

RACHNA with a dynamic threshold (RACHNA$_{dt}$) eliminates this restriction by using the increment $\epsilon_i=1/n$, so that the $\lvert S_{t_j}\rvert\times\epsilon_i > u_j$ threshold is determined dynamically by the problem input, rather than being a fixed parameter. A dynamic threshold permits coalitions of all sizes $\leq n$ \cite{Diehl2022ScalableCollectives}. RACHNA$_{dt}$'s computational and communication complexities are $O(mn^3\log n (u_{max}))$ and $O(n^3m\lvert S \rvert(u_{max}))$, respectively.

RACHNA$_{dt}$ is decentralized, permits multiple services, and produces empirically optimal solutions with up to 1000-robot collectives and 100-robot coalitions \cite{Diehl2022ScalableCollectives}. However, RACHNA$_{dt}$ does not fully satisfy near real-time domains' requirements. RACHNA$_{dt}$'s runtimes are long (i.e., $>$ 2 hours) when the services per robot to overall service types ratio is high (e.g., homogeneous collectives with multiple services). Additionally, RACHNA$_{dt}$ can require excessive total communication (i.e., $>27$ GB). Even given these limitations, RACHNA$_{dt}$ is one of only two algorithms demonstrated to provide high-quality task allocations for multiple service collectives with unrestricted coalition sizes \cite{Diehl2022ScalableCollectives}. As such, RACHNA$_{dt}$ is considered as a baseline in this manuscript.

\subsubsection{Simultaneous Descending Auction (SDA)} SDA incorporates a \textit{descending auction} protocol with the same mapping of task and service agents to buyers and sellers as RACHNA and RACHNA$_{dt}$ \cite{Service2014AAllocation}. Each robots' wage is initially $u_{max}+\epsilon_d$, where $\epsilon_d$ is a fixed wage decrement. Robot's wages are decremented at the beginning of each bidding round according to a wage decrement schedule (e.g., decrement the robot with the highest wage), and task agents that still require additional services bid on service bundles. Tasks determine which robots to bid on using bipartite matching. A task agent is awarded a bundle if the bid is at least the robots' current wages. The auction stops when all robots have been purchased, or all wages are zero. SDA's computational complexity is $O(mn^4u_{max}/\epsilon_d$) \cite{Service2014AAllocation}, and its communication complexity is $O((m+n)\lvert S\rvert u_{max}/\epsilon_d)$ \cite{Labella2006DivisionBehavior}.

SDA is decentralized and permits multiple service types. Additionally, SDA produces near-optimal solutions with $\leq$ 1000 robots and $\leq100$ robot coalitions, while using $<600$ MB total communication \cite{Diehl2022ScalableCollectives,Service2014AAllocation}.  However, SDA's runtimes can be longer than near real-time (i.e., $>25$ min)  \cite{Diehl2022ScalableCollectives}. SDA is considered as a baseline, despite this limitation, as it is the auction-based algorithm that comes closest to satisfying the near real-time domains' requirements \cite{Diehl2022ScalableCollectives}.

\subsection{Hedonic Game-Based Coalition Formation}

Some recent coalition formation algorithms leverage hedonic games. A \textit{hedonic game} is a non-cooperative, game theoretic coalition formation model, in which robots prefer certain coalitions \cite{Monaco2019OnWelfare}. A problem instance comprises a set of $n$ robots, $A = \{a_1, \dots, a_n\}$, and  and a preference profile, $\succsim=(\succsim_1, \dots, \succsim_n)$, where robot $a_i$ prefers coalition $C_1$ to coalition $C_2$, if $C_1 \succsim_i C_2$ \cite{Aziz2019FractionalGames,Aziz2016HedonicGames}.  A robot will leave its coalition for a preferable one, if such a coalition exists. Coalition structures in which no robots prefer other coalitions are \textit{Nash stable} \cite{Aziz2016HedonicGames} and only exist for some preference profiles \cite{Jang2018AnonymousSystem}. Hedonic games are well-studied (e.g., \cite{Carosi2019LocalGames,Dreze1980HedonicStability,Fichtenberger2021TestingGames,Monaco2018StableGames,Nguyen2016AltruisticGames}); however, hedonic game algorithms do not transfer directly to robot coalition formation for task allocation, as low-utility, stable solutions can exist. A key challenge in leveraging hedonic games is designing robots’ individual preferences to correspond to globally appropriate solutions.

Existing hedonic game-based coalition formation incorporates \textit{anonymous} hedonic games, where robots' preferences depend only on coalition size \cite{Czarnecki2019HedonicRobots,Czarnecki2021ScalableRobots,Jang2018AnonymousSystem}. Anonymous hedonic games in which robots prefer smaller coalitions are guaranteed Nash stable solutions \cite{Jang2018AnonymousSystem}. A centralized anonymous hedonic game produces near-optimal solutions in real-time (i.e., $<2$ seconds) with up to 2000 robots; however, centralization is incompatible with deployed networks \cite{Czarnecki2019HedonicRobots,Czarnecki2021ScalableRobots}. The family of distributed anonymous algorithms, called GRoup Agent Partitioning and Placing Event algorithms, are more applicable \cite{Dutta2021DistributedAllocation,Jang2018AnonymousSystem, Jang2019AnRequirements}.

\subsubsection{GRoup Agent Partitioning and Placing Event (GRAPE)}

GRAPE was originally designed for homogeneous, single-service collectives. The preference profile is defined by a task-based peaked reward, which is highest when a robot's coalition possesses exactly the robots required for its task, divided by the coalition size (Equation \ref{eq:peaked_reward}) \cite{Jang2018AnonymousSystem}. An individual robot's reward for coalition $C_j$ and task $t_j$, where $t_j$ has utility $u_j$ and requires $k$ robots is:

\begin{equation}
    \label{eq:peaked_reward}
    \frac{u_j}{k} \times e^{-\frac{\lvert C_j \rvert}{k}+1}.
\end{equation} 

\noindent This reward decreases as a coalition grows, so a Nash stable solution exists \cite{Jang2018AnonymousSystem}. The computational complexity is $O(n^2md_G)$ with a communication complexity of $O(n^3d_G)$, where $d_G$ is the communication network topology's diameter \cite{Jang2018AnonymousSystem}.

Homogeneous GRAPE uses local communication, and produces empirically optimal solutions in near real-time (i.e., $<5$ min) for single-service collectives with a centralized simulator \cite{Diehl2022ScalableCollectives}; however, it does not address heterogeneous collectives or homogeneous collectives with multiple service types.

GRAPE was extended to allow for differences in service performance quality \cite{Jang2019AnRequirements}, and the most recent version permits different resources (e.g., sensors and actuators) \cite{Dutta2021DistributedAllocation}. However, no prior variant is compatible service model.

\section{GRAPE with Services Extensions}

This manuscript proposes two algorithms: GRAPE with the Services model (GRAPE-S) and Pairwise GRAPE-S (Pair-GRAPE-S). GRAPE-S integrates GRAPE with a services model (Algorithm \ref{alg:grapes}). Robots are initially assigned to a \textit{void task}, $t_\emptyset$ (line 2).  Robot $a_i$ selects its preferred coalition (lines 4-9) at the start of each algorithm iteration, $r^i$, broadcasts its beliefs, $\Pi^i$, about all robots' current task assignments (line 10), and updates its belief states based on messages from neighboring robots, $N_i$, in the network topology (lines 11-18). A robot's message is given precedence if the robot's belief state has been updated more times or more recently than the receiving robot's belief state. The robot id is used as a tiebreaker as needed to avoid dividing the collective. This precedence system serves as a \textit{distributed mutex}, which allows only a single robot to alter the valid coalition assignments during each iteration \cite{Jang2018AnonymousSystem}. The algorithm finishes when Nash stability is reached.

\begin{algorithm}
\caption{GRAPE-S on agent $a_i$. \textcolor{blue}{Blue text} denotes the lines that differ from GRAPE \cite{Jang2018AnonymousSystem}}\label{alg:grapes}
\tcc{Initialization}
 $r^i \gets 0$\tcp*{partition update counter} 
\textcolor{blue}{$\pi^i \gets \{C_{\emptyset} = \{(a_i, \emptyset)\}\forall a_i \in A \}, C_j = \emptyset \forall t_j \in T\}$}\;
 \While{solution is not Nash Stable}{
 \tcc{Select highest utility task/service}
   \textcolor{blue}{ $(t_{j*}, C_{j*}, s_{*}) \gets \argmax_{\forall C_j \in \pi^i, \forall s_{l} \in S}$ utility$(t_j, C_j \cup \{(a_i, s_{l})\})$}\;
 \If{\textcolor{blue}{utility$(t_{j*}, C_{j*}) > $utility$(t_{\pi^i(i)}, C_{\pi^i(i)})$}}
  {
    \textcolor{blue}{Join $C_{j*}$ performing service $ s_{*}$ and update $\Pi^i$}\;
    $r^i \gets r^i + 1$\;
    $time^i \gets$ current time\;
    }
 
\tcc{Broadcast local partition to neighbors}
Broadcast $M^i = \{r^i, time^i, \Pi^i\}$ and receive $M^l$ from neighbors $\forall a_l \in N_i$\;
Construct $M^i_{rcv} = \{M^i, \forall M^l\}$\;
\tcc{Distributed Mutex}
\For{each message $M_l\in M^i_{rcv}$}{
\If{$(r^i < r^k)$ or $(r^i=r^k$ $\&$ $time^i<time^k)$}{
    $r^i \gets r^k$\;
    $time^i \gets time^k$\;
    $\Pi^i \gets \Pi^k$\;
}
}
}
\end{algorithm}

The key difference between GRAPE-S and GRAPE is how robots select coalitions (lines 4-9). Homogeneous GRAPE's peaked reward is task-based, while GRAPE-S uses a novel peaked reward, in which robots prefer coalitions based on tasks and services. An individual robot's reward for joining a coalition to perform service $s$ for task $t_j$ is:
 
 \begin{equation}
    \label{eq:peaked_reward_new}
    \frac{u_{js}}{\lvert S_s\rvert} \times e^{-\frac{\lvert C_s \rvert}{\lvert S_s\rvert}+1},
\end{equation} 

\noindent where $\lvert S_s\rvert$ is the number of robots $t_j$ requires to perform service $s$, $C_s$ is the set of robots assigned to perform $s$, and $u_{js}$ is the utility of assigning $t_j$ sufficient robots to perform $s$. This reward enables GRAPE-S to produce coalition formation solutions with multiple service types, while using $O(n^2\lvert S \rvert md_G)$ computation and $O(n^3d_G)$ communication \cite{Diehl2022ScalableCollectives}.

This manuscript considers $u_{js}=\lvert S_s \rvert$, so allocating a robot to perform any service is equally valuable. This reward model is reasonable when there are sufficient robots to perform all tasks, because it encourages robots to provide services to the tasks that are farthest from meeting their service requirements. 

\subsection{Pairwise GRAPE-S (Pair-GRAPE-S)}

A risk of extending GRAPE-S to heterogeneous, multiple service collectives is that it becomes possible for the algorithm to make \textit{suboptimal assignments} that block an optimal solution from being found. For example, consider a robot assigned to $t_1$, which must be assigned to $t_2$ for a solution to be optimal. Additionally, suppose that switching to $t_2$ is only beneficial if another robot assumes responsibility for $t_1$ (e.g., $t_1$ is more mission-critical than $t_2$, so the reward for meeting its service requirement is higher). Only individually beneficial coalition changes occur in GRAPE-S, with the goal of Nash stability, making the optimal solution unreachable.

An additional extension mitigates this risk by having the collective achieve pairwise stability (i.e., no two robots can benefit by individually or mutually changing their coalition membership) after initial Nash stability is reached. GRAPE-S plus pairwise stability is denoted as Pair-GRAPE-S (Algorithm \ref{alg:pairwise}).

An agent $a_o$ assesses if the pairwise stability step is needed by determining whether there are tasks assigned insufficient robots, and there are unassigned robots (lines 2-3). This implementation limits a task's coalition size to the number of robots required; however, the reduction from an implementation that permits robots exceeding a task's requirements is straightforward, requiring only a preprocessing step that unassigns the excess robots. 


\begin{algorithm}
\caption{Pairwise stability step: Pairwise Stability (Pair-GRAPE-s) on agent $a_o$}
\label{alg:pairwise}
$\Pi \gets$ GRAPE-S($\Pi$, $T$)\;
\While{$\exists t_j\in T$ \textnormal{s.t.} $t_j$ \textnormal{is assigned insufficient robots in} $\Pi$ \textnormal{\\and} $\exists$ $a_i \in \Pi$ \textnormal{assigned to} $t_\emptyset$ }{
\If{$a_o$ \textnormal{assigned} $t_\emptyset$} {
  \textnormal{Identify a robot $a_i$ allocated a task/service} $a_o$ \textnormal{can perform}\;
  \textnormal{Communicate request to assume $a_i$'s task/service requirement}\;
  \If{\textnormal{request accepted}} {
    \textnormal{Reassign $a_o$ to $a_i$'s task/service requirement}\;
    \textnormal{Assign $a_i$ to the unmet task/service requirement it requested}\;
    \textnormal{Broadcast new belief state};
  }
  
}
 \textnormal{Determine new belief state with the distributed mutex}\;
}
\end{algorithm}

If the pairwise stability step is necessary, and $a_o$ is itself unassigned, $a_o$ seeks a robot $a_i$ that can benefit from mutually changing coalitions (i.e., an $a_i$ that can perform an unmet task requirement and is currently assigned a service that $a_o$ can perform) (lines 4-12). When such an $a_i$ is identified, $a_o$ assumes $a_i$'s current assignment and reassigns $a_i$ to the unmet task requirement with the highest reward (lines 8-9). The new belief state is broadcast (line 10), and the distributed mutex updates the belief state based on neighbors' messages (line 13, Alg. 1 lines 11-19).

\begin{lemma}
\label{lem:1}
    Algorithm 2 produces a Nash stable solution, given the reward function specified by Eq. \ref{eq:peaked_reward_new} and $u_{js}=\lvert S_s \rvert$.
\end{lemma}

\begin{proof}
    Let $\Pi$ be the Nash stable partition produced by GRAPE-S. Additionally, consider a single mutual coalition change made by agents $a_1$ and $a_2$, where $a_1$ was initially unassigned. The resulting partition is denoted as $\Pi'$.

    All coalitions in $\Pi$ and $\Pi'$ are the same size, except for $a_2$'s new coalition in $\Pi'$, which gained one agent. The reward for coalition membership depends only on the coalition size; thus, every coalition in $\Pi'$ has an associated reward that is less than or equal to its reward in $\Pi$. Agents $A\setminus a_2$ cannot benefit from individually changing coalitions in $\Pi'$, as there was no benefit in $\Pi$, where the reward was potentially higher. These agents meet the requirement for Nash stability.

    Agent $a_2$ was assigned the unsatisfied task requirement with the highest reward. Additionally, Eq. 4.2 with $u_{js}=\lvert S_s \rvert$ sets the reward for every unsatisfied requirement higher than every satisfied requirement; thus, $a_2$ cannot benefit from individually changing its coalition membership, satisfying Nash stability. As each of Algorithm 2's changes to a Nash stable partition produces another Nash stable partition, the final result will also be Nash stable.

\end{proof}

\begin{theorem}
\label{proof:t1}
Algorithm 2 produces a pairwise stable solution, given the reward function specified by Eq. 4.2 and $u_{js}=\lvert S_s \rvert$.
\end{theorem}

\begin{proof}

    Lemma \ref{lem:1} satisfies the individual stability requirement. It remains to show that no two agents can benefit from mutually changing coalitions.
    
   Consider two arbitrary agents, $a_1$ and $a_2$. If both are unassigned, there is no benefit to mutually changing coalitions. The agents can only lower each other's potential reward (i.e., by joining the same coalition), not increase it, and changing coalitions individually already provides no benefit, as the partition is Nash stable.

   Similarly, suppose that both agents are assigned tasks. Exchanging coalitions cannot be mutually beneficial, as it merely exchanges the robots' rewards, unaltered. Additionally, the agents can only lower each other's potential reward by joining the same coalition, and changing coalitions individually provides no benefit. 
   
   Finally, suppose that only $a_1$ is assigned a task. There is no benefit in $a_1$ and $a_2$ joining the same or separate coalitions, due to the preference for smaller coalitions and the Nash stable partition. Agent $a_2$ does benefit from replacing $a_1$; however, $a_1$'s reward decreases if it joins a coalition that satisfies a task's requirements, and $a_1$ cannot join a coalition with unsatisfied requirements, as Algorithm 2 already made all beneficial exchanges; thus, Algorithm 2's solution is pairwise stable.
 \end{proof}

Algorithm 2 requires at most $O(n)$ reassignments, although it is expected to be much lower in practice. Each reassignment requires a robot to communicate with up to $O(n)$ other robots, each of which performs at most $O(m\lvert S\rvert)$ computations to determine the unmet service requirement with the highest reward. The resulting computational complexity is $O(n^2\lvert S\rvert m)$ with $O(n^2)$ communication. These complexities are lower than GRAPE-S's, meaning that Pair-GRAPE-S's overall computational and communication complexities are identical to GRAPE-S. Centralized and distributed evaluations assessed GRAPE-S's and Pair-GRAPE-S's suitability for near real-time domains.

\section{Centralized Evaluation}

A centralized simulation-based evaluation provided an initial assessment of GRAPE-S's and Pair-GRAPE-S's ability to produce collective coalition formation solutions for near real-time domains, compared to the two auction baselines: SDA and RACHNA$_{dt}$. An algorithm is considered \textit{viable} for near real-time domains if it produced near-optimal solutions (i.e., $>95$\% utility) in near real-time (i.e., $<5$ min). Communication was also assessed, as minimal distributed communication can be preferable when permanent communication infrastructure is unavailable or limited.

The evaluation considered \textit{achievable missions}, meaning that there were sufficient robots to perform all tasks. Real-world missions will ideally be achievable; however, collectives are unlikely to possess substantially more robots than necessary, due to the expense and logistical challenges \cite{Diehl2022ScalableCollectives}. Each considered mission required exactly the number of robots in the collective. 

Collectives with 100, 500, and 1000 robots were considered (Table \ref{table:ind_variables_achievable}). The \textit{percent tasks} variable determined the numbers of tasks (i.e., 1\%, 10\%, and 50\% the collective size) and the average coalition sizes (i.e., 100, 10, and 2 for 1\%, 10\%, and 50\% tasks, respectively). Large-scale tasks (e.g., clearing an urban area of hazards), which require larger coalition sizes with many services, map to 1\% tasks. Medium-scale tasks (e.g., assessing a damaged building) map to 10\% tasks, and smaller-scale tasks (e.g., guarding a building entrance) map to 50\%.  1\% tasks (i.e., coalition size 100) was used only for collectives with $>100$ robots in order to have more than one task per mission. Four \textit{collective compositions} were considered, each with a different combination of overall number of \textit{service types} and \textit{services per robot}. More capable robots have more services, while the number of different service types and services per robot combinations determines the degree of heterogeneity. More heterogeneous problems are more difficult, because fewer robots are interchangeable.

\begin{table}[h]
    \centering
    \normalsize
    \caption{Independent variables}
    \begin{tabular}{|c|l|}
        \hline
        Collective Size & 100, 500, 1000 \\
        \hline
        Percent Tasks & 1, 10, 50\\
        \hline
        Service Types: Per Robot & 5:1 , 5:5, 10:1, 10:5\\
        \hline
    \end{tabular}
    \label{table:ind_variables_achievable}
  
\end{table}

Pair-GRAPE-S was considered only for collectives with ten service types and five services per robots. Pair-GRAPE-S is identical to GRAPE-S for the other three collective compositions, where the collectives are homogeneous or possesses only single service robots.

Twenty-five problem instances were randomly generated per independent variable combination, for a total of 800 trials per algorithm.  Robots' and tasks' services were selected randomly, and each task was assigned a random utility in the range [1, 50], where higher utility tasks are more important to the mission. Trials were performed on a HP Z640 Workstation (Intel Xeon processor, 62 GB RAM) using a centralized C++ simulator \cite{Sen2013AFormation}, which performed each algorithm iteration for each robot sequentially. The simulator used a fully-connected communication topology and assumed instantaneous communication.

The dependent variable \textit{runtime} is the time in minutes (min) and seconds (s) required for an algorithm to produce a solution. \textit{Total communication} is the sum of all message sizes in megabytes (MB). \textit{Percent utility} is the solution utility as a percentage of the total possible mission utility (i.e., the solution utility divided by the sum of all the task utilities). Overall, high percent utilities with low runtimes and low total communication are preferred. GRAPE-S were \textit{hypothesized} to have better overall performance than the baseline algorithms SDA and RACHNA$_{dt}$. Specifically, GRAPE-S and Pair-GRAPE-S were hypothesized to require shorter runtimes ($H_1$) and less communication ($H_2$) without substantially lower percent utilities ($H_3)$. Pair-GRAPE-S's utilities were additionally hypothesized to be higher than GRAPE-S's for ten service collectives with five services per robot ($H_4$).

\subsection{Results}

GRAPE-S, SDA, and RACHNA$_{dt}$ produced solutions for all 800 trials. Pair-GRAPE-S also produced solutions for all of its 200 ten service types/five services per robot trials. Recall that Pair-GRAPE-S uses its extra pairwise stability step only when a suboptimal solution is initially produced. The pairwise stability step ran during 111 trials. For the other 89 trials, the pairwise stability was unnecessary, and Pair-GRAPE-S ran the equivalent of GRAPE-S. The results are presented by dependent variable.  Box plots and non-parametric statistical analyses are used, as the results were not normally distributed. 

GRAPE-S's results are presented first for each dependent variable. Kruskal-Wallis analysis with Mann-Whitney-Wilcoxon post-hoc tests assessed if GRAPE-S's results changed significantly with respect to the collective sizes and percent tasks. The collective size analysis considered only 10\%-50\% tasks, and the percent task analysis considered only 500-1000 robots, as 100 robots with 1\% tasks were not evaluated. GRAPE-S's results with five and ten service types, as well as one and five services per robot were compared using Mann-Whitney-Wilcoxon tests. 

Pair-GRAPE-S's results are presented next. Like GRAPE-S, Pair-GRAPE-S's performance was compared across collective sizes and percent tasks using Kruskal-Wallis analysis with Mann-Whitney-Wilcoxon post-hoc tests. The number of services and services per robot were not considered, as Pair-GRAPE-S was applicable only for one collective composition.

Each results section concludes by comparing GRAPE-S and Pair-GRAPE-S to the auction-based baselines. Friedman tests with Conover post-hoc analyses assessed whether GRAPE-S's results differed significantly from the auction baseline algorithms. This analysis considered all trials, across all collective compositions, collective sizes, and percent tasks. Additional Friedman tests with Connover post-hoc analysis assessed whether Pair-GRAPE-S's results differed significantly from GRAPE-S and the auction baselines. This analysis considered only collectives with ten service types and five services per robot. Note that statistical analysis of the auction baselines individually is not included, as this analysis has been published previously \cite{Diehl2022ScalableCollectives}.

\subsubsection{Runtime Results}

\begin{figure*}[h!]
    \subfloat[Five service types, one service per robot. \label{subfig:runtime1}]{%
      \includegraphics[width=0.49\linewidth]{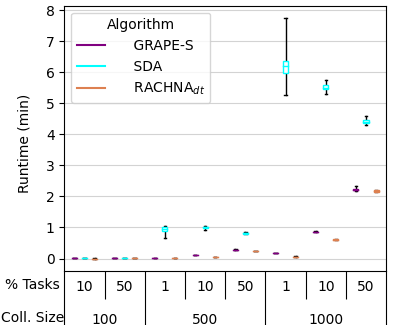}
    } 
    \hfill
    \subfloat[Five service types, five services per robot. \label{subfig:runtime2}]{%
      \includegraphics[width=0.49\linewidth]{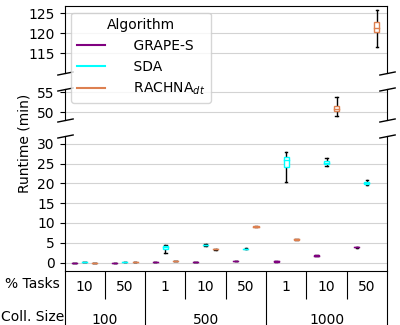}
    } 
    \hfill
     \subfloat[Ten service types, one service per robot. \label{subfig:runtime3}]{%
      \includegraphics[width=0.49\linewidth]{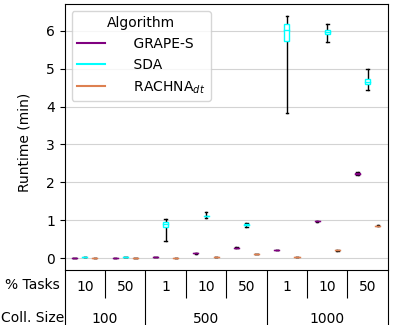}
    } 
    \hfill
    \subfloat[Ten service types, five services per robot. \label{subfig:runtime4}]{%
      \includegraphics[width=0.49\linewidth]{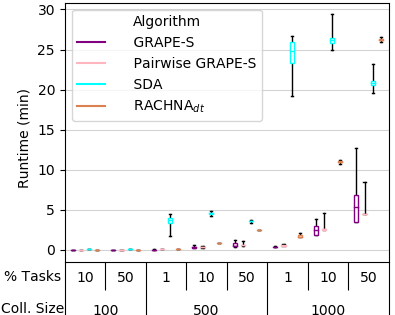}
    } 
     \caption{Centralized algorithm runtimes by collective size, percent tasks, services, and services per robot. Pair-GRAPE-S is included only in Figure \ref{subfig:runtime4}, as it is otherwise identical to GRAPE-S. Each subfigure has a different y-axis maximum. Figures \ref{subfig:runtime1} and \ref{subfig:runtime3} have 1 min increments, while \ref{subfig:runtime2} and \ref{subfig:runtime4} have 5 min increments. Plot \ref{subfig:runtime3} incorporates axis breaks to help readability}
    \label{fig:runtimes_achievable}
\end{figure*}

GRAPE-S's runtimes were well within the 5 min target for most independent variable combinations (see Figure \ref{fig:runtimes_achievable}). The only exception was 1000 robot collectives with 50\% tasks, ten service types, and five services per robot (median 5 min 22 s). The the maximum runtime was 12 min 15 s, but all other trials required $\leq7$ min 18 s, which is reasonable for near real-time domains. GRAPE-S's runtimes did increase significantly with increased collective size ($H$ ($n=200$) $=$ 523.03, $p < 0.01$) and percent tasks ($H$ ($n=200$) $=$ 271.95, $p < 0.01$), consistent with GRAPE-S's $O(n^2\lvert S\rvert m)$ computational complexity. All post-hoc analyses were also significant ($p<0.01$); however, runtimes remained low. GRAPE-S's runtimes with five services types and one service per robot (Figure \ref{subfig:runtime1}) were shorter than those with five service types and five services per robot (Figure \ref{subfig:runtime2}); however, they differed from those with ten service types and one service per robot by $<10$ s (see Figure \ref{subfig:runtime3}). Runtimes with ten service types and five services per robot (Figure \ref{subfig:runtime4}) were much longer and more variable, which occurred due to the number of iterations required to produce a solution. All other collective compositions required exactly $n$ iterations, where $n$ is the collective size, while GRAPE-S with ten service types and five services per robot frequently required more. GRAPE-S's runtimes differed significantly from one to five services per robot ($p<0.01$). No significant difference was found between five and ten service types.

Pair-GRAPE-S's runtimes were also well within the 5 min target for most independent variable combinations (Figure \ref{subfig:runtime4}). The only trial that exceeded 5 min occurred with 1000 robots and 50\% tasks, which required 8 min 17 s. Like GRAPE-S, Pair-GRAPE-S's runtimes increased significantly with collective size ($H$ ($n=50$) $=$ 132.46, $p < 0.01$), and percent tasks ($H$ ($n=50$) $=$ 65.29, $p < 0.01$), and all post hoc analyses were significant ($p<0.01$). Overall, Pair-GRAPE-S's runtimes were generally less variable than GRAPE-S's.

\paragraph{Baseline Comparison:} Overall, GRAPE-S's and Pair-GRAPE-S's runtimes were better than SDA's and RACHNA$_{dt}$'s. However, differences between the algorithms with 100 robots were negligible, and no one algorithm was the fastest for all independent variable combinations (Figure \ref{fig:runtimes_achievable}). 

RACHNA$_{dt}$, like GRAPE-S, completed all trials in $<5$ min with five service types and one service per robot (Figure \ref{subfig:runtime1}). SDA's runtimes for these variable values were $<8$ min, exceeding the 5 min threshold, but all of these results are reasonable for near real-time coalition formation. 

GRAPE-S was the fastest algorithm with five services and five services per robot (Figure \ref{subfig:runtime2}). All algorithms' runtimes for this collective composition were reasonably near the 5 min target with 500 robots; however, GRAPE-S's $<5$ min runtimes with 1000 robot collectives were substantially better than SDA's 20-30 min runtimes. RACHNA$_{dt}$'s runtimes were $<5.5$ min with 1\% tasks, but increased to 45-55 min with 10\% tasks, and were $>2$ hours with 50\% tasks, substantially exceeding the $5$ min target. All algorithms' performance with ten services and one service per robot (Figure \ref{subfig:runtime3}) was similar to five services and one service per robot. The only notable difference was that RACHNA$_{dt}$ was faster, due to the lower services per robot to service types ratio \cite{Diehl2022ScalableCollectives}.

Pair-GRAPE-S was the fastest with ten services and five services per robot, closely followed by GRAPE-S (Figure \ref{subfig:runtime4}). RACHNA$_{dt}$ generally performed comparably, completing most trial in $<5$ min; however, Pair-GRAPE-S and GRAPE-S  with 1000 robots and 10\%- 50\% tasks were meaningfully faster (i.e., $\approx$ 2x). Pair-GRAPE-S and GRAPE-S with 500 robots were also marginally faster than SDA, with $<5$ min runtimes; however, Pair-GRAPE-S's $\leq 8$ min 12 s runtimes, and GRAPE-S's $\leq12$ min 15 s runtimes with 1000 robot collectives were substantially faster than SDA's 20-30 min runtimes.

The algorithms’ performance with ten services and five services per robot differed significantly ($\chi_r^2$ ($n=200$) $=$ 484.15, $p<0.01$). Post hoc analysis found significant differences between Pair-GRAPE-S and all other algorithms (GRAPE-S/Pair-GRAPE-S: $p=0.03$, all others: $p<0.01$), as well as GRAPE-S and all other algorithms (GRAPE-S/Pair-GRAPE-S: $p=0.03$, all others: $p<0.01$). No significant difference between SDA and RACHNA$_{dt}$ was identified. A significant difference was also identified across GRAPE-S, SDA, and RACHNA$_{dt}$ ($\chi_r^2$ ($n=800$) $=$ 913.24, $p<0.01$), considering all collective compositions. Specifically, significant differences existed between pairs: GRAPE-S/SDA ($p<0.01$), GRAPE-S/RACHNA$_{dt}$ ($p=0.02$), and SDA/RACHNA$_{dt}$ ($p<0.01$).

\subsubsection{Communication Results}

\label{sec:comms}

\begin{figure*}[t]
    \subfloat[Five service types, one service per robot. \label{subfig:comms1}]{%
      \includegraphics[width=0.49\linewidth]{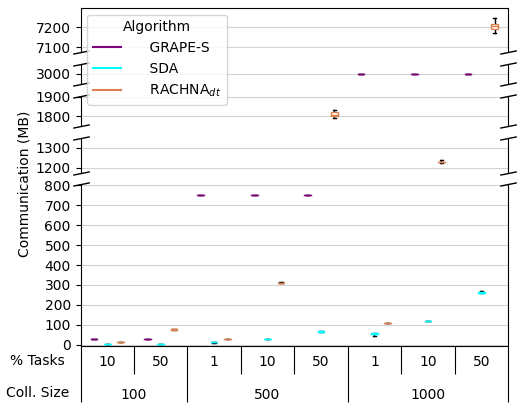}
    } 
    \hfill
    \subfloat[Five service types, five services per robot. \label{subfig:comms2}]{%
      \includegraphics[width=0.49\linewidth]{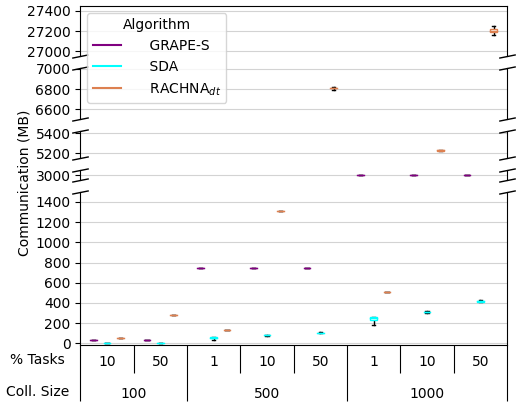}
    } 
    \hfill
     \subfloat[Ten service types, one service per robot. \label{subfig:comms3}]{%
      \includegraphics[width=0.49\linewidth]{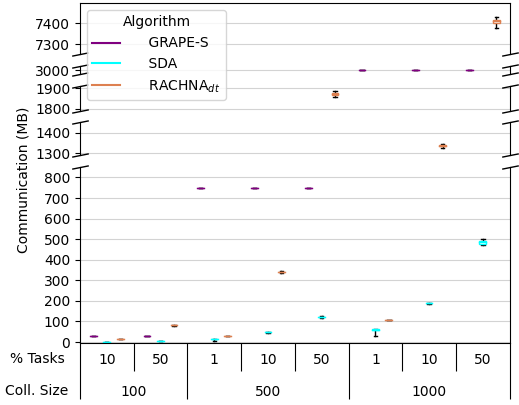}
    } 
    \hfill
    \subfloat[Ten service types, five services per robot. \label{subfig:comms4}]{%
      \includegraphics[width=0.49\linewidth]{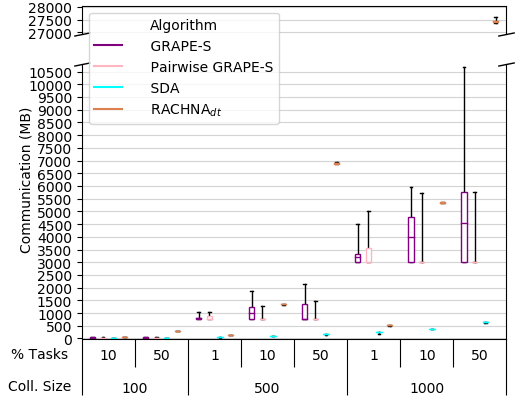}
    } 
   \caption{Centralized algorithms' total communication by collective size, percent tasks, number of services, and number of services per robot.  Pair-GRAPE-S is included only in Figure \ref{subfig:comms4}, as it is otherwise identical to GRAPE-S. Each subfigure has a different y-axis maximum. Subfigures \ref{subfig:comms1} and \ref{subfig:comms3} have 100 MB increments. Subfigures \ref{subfig:comms2} and \ref{subfig:comms4} have 200 MB and 500 MB increments, respectively. All plots include axis breaks to help readability}
    \label{fig:comms_achievable}
\end{figure*}

GRAPE-S's communication requirement was identical for collectives with five service types and one service per robot (Figure \ref{subfig:comms1}), five service types and five services per robot (Figure \ref{subfig:comms2}), and ten service types and one service per robot (Figure \ref{subfig:comms3}). The communication for these collective compositions depended only on the collective size and was not impacted by the percent tasks or random differences between trial inputs, as each trial required exactly $n$ iterations. The required communication was low with 100 robots (i.e., 30 MB), but increased substantially with increased collective size, consistent with GRAPE-S's $O(n^3)$ communication complexity. GRAPE-S required 750 MB with 500 robots, and 3,000 MB (i.e., 3 GB) with 1000 robots. 

GRAPE-S with ten service types and five services per robot (Figure \ref{subfig:comms4}) required at minimum the same communication as the other collective compositions; however, the communication ranges increased with increased collective size and increased percent tasks, resulting in much higher worst case communication (i.e., $>10,000$ MB, 10 GB). Trials requiring exactly $n$ iterations used the least communication, while trials requiring $>n$ iterations used more.

Overall, GRAPE-S's communication increased significantly across the collective sizes ($H$ ($n=200$) $=$ 566.54, $p<0.01$) with significant post-hoc analysis results ($p<0.01$). Significant differences were also detected between results for five and ten service types ($p=0.02)$, and one and five services per robot ($p=0.02$), due to differences between collectives with ten service types and five services per robot and all other considered collective compositions. No significant difference was identified with respect to the percent tasks.

Pair-GRAPE-S required a similar amount of communication as GRAPE-S for collectives with ten services and five services per robot. Like GRAPE-S, Pair-GRAPE-S's total communication increased significantly across the collective sizes ($H$ ($n=50$) $=$ 133.39, $p<0.01$) with significant post hoc analyses ($p<0.01$), and no significant difference was detected across the percent tasks; however, Pair-GRAPE-S's communication was slightly more variable than GRAPE-S's for 1\% tasks, and much less variable for 10\%-50\% tasks. The decreased variability with 10\%-50\% tasks resulted in lower overall communication, as Pair-GRAPE-S's results were concentrated at the lower end of GRAPE-S's range.

\paragraph{Baseline Comparison}

  SDA required the least communication overall (Figure \ref{fig:comms_achievable}). GRAPE-S and Pair-GRAPE-S each required the second least communication for certain independent variable combinations, while RACHNA$_{dt}$ required the second least for others. All algorithms with 100 robots required little communication, but had high communication requirements for certain 1000 robot trials.

GRAPE-S required more communication than RACHNA$_{dt}$ with five service types, one service per robot, and 1-10\% tasks, but less with 50\% tasks (Figure \ref{subfig:comms1}). Overall, RACHNA$_{dt}$ required little communication with 100 robots (i.e., $<100$ MB), or with 500 robots and 1\%-10\% tasks (i.e., $<400$ MB). However, RACHNA$_{dt}$ required $>1,800$ MB (i.e., 1.8 GB) with 500 robots and 50\% tasks. Additionally, RACHNA$_{dt}$ required $<200$ MB with 1000 robots and 1\% tasks, but $>1,200$ MB (i.e., 1.2 GB) with 10\%-50\% tasks. SDA required $<500$ MB communication for all trials, which is relatively low.

GRAPE-S required more communication than RACHNA$_{dt}$ for most trials with five services, five services per robot, and 1\%-10\% tasks (Figure \ref{subfig:comms2}). 500 robots and 10\% tasks were an exception, where RACHNA$_{dt}$ required more communication than GRAPE-S (i.e., $>1,200$ MB, or 1.2 GB). RACHNA$_{dt}$ also required more communication than GRAPE-S with 50\% tasks. RACHNA$_{dt}$ required $<500$ MB, except with 500 robots and 50\% tasks (i.e., $>6,800$ MB, 6.8 GB), and 1000 robots with 50\% tasks (i.e., $>5,200$ MB, 5.2 GB). SDA required $<550$ MB for all trials. All algorithms' performance with ten services and one service per robot (Figure \ref{subfig:comms3}) was similar to five services and one service per robot. The only notable difference was that SDA and RACHNA$_{dt}$ required slightly more communication.

The communication for collectives with ten services and five services per robot also depended on the percent tasks  (Figure \ref{subfig:comms4}). GRAPE-S and Pair-GRAPE-S required less communication than RACHNA$_{dt}$ for all 50\% task trials, as well as some 10\% task trials; however, RACHNA$_{dt}$ required less communication for other 10\% trials, due to GRAPE-S's and Pair-GRAPE-S's high variability. Finally, Pair-GRAPE-S and GRAPE-S with 1\% tasks required more communication than RACHNA$_{dt}$. RACHNA$_{dt}$'s required communication was relatively low with 1\% tasks (i.e., $<600$ MB), but was $>1,000$ MB (i.e., 1 GB) with 10\%-50\% tasks. RACHNA$_{dt}$ required $>27,500$ MB (i.e., 27.5 GB) in the worst-case. SDA required the least communication, $<500$ MB for most trials, with a  $>650$ MB worst-case communication requirement.

The algorithms' communication with ten services and five services per robot differed significantly ($\chi_r^2$ ($n=200$) $=$ 399.34, $p<0.01$). No significant difference between Pair-GRAPE-S and GRAPE-S was found, but all other post hoc tests were significant ($p<0.01$). The algorithms also differed significantly overall ($\chi_r^2$ ($n=800$) $=$ 1186.8, $p<0.01$). Significant differences existed between SDA and GRAPE-S ($p<0.01$), and SDA and RACHNA$_{dt}$ ($p<0.01$), but no significant difference between GRAPE-S and RACHNA$_{dt}$ was identified.


\subsubsection{Utility Results}

GRAPE-S produced optimal solutions with five services and one service per robot, five services and five services per robot, and ten services and one service per robot (see Table \ref{table:utilities_achievable}). GRAPE-S's percent utilities with ten services and five services per robot were slightly lower, but still near-optimal in most cases. 

\begin{table}[h]
\normalsize
    \centering
        \caption{GRAPE-S's percent utility statistics with ten service types and five services per robot. 1\% tasks had 100\% utility, as did all other collective compositions}
    \begin{tabular}{|c|c|c|}
    \hline
          Percent Tasks & Collective Size & Median (Minimum, Maximum)\\
         \hline
         \multirow{3}{*}{10} & 100 & 100.0 (89.68, 100.0)\\
         & 500 & 100.0 (98.47, 100.0) \\
         & 1000 & 100.0 (100.0, 100.0)\\
         \hline
          \multirow{3}{*}{50}&  100 & 100.0 (96.59, 100.0)\\
         & 500 & 100.0 (99.66, 100.0)\\
         & 1000 & 100.0 (99.69, 100.0)\\
         \hline
    \end{tabular}
    \label{table:utilities_achievable}
\end{table}

GRAPE-S's lowest utility trial (i.e., 89.68\%), which occurred for 100 robot collectives with ten services, five service per robot, and 10\% tasks, was unique in its low solution quality. The next lowest percent utility with this independent variable combination was 96.07\%. The percent utilities for all other independent variable combinations were above the $>95$\% target threshold.

GRAPE-S's percent utilities differed significantly across the percent tasks ($H$ ($n=200$) $=$ 7.19, $p=0.03$), with significant differences existed between 1\% and 50\% tasks ($p=0.01)$. No other significant differences between percent tasks were detected. However, significant differences were found between five and ten services ($p<0.01$), and between one and ten services per robot ($p<0.01$). Pair-GRAPE-S produced optimal solutions for all problems.

\paragraph{Baseline Comparison}

RACHNA$_{dt}$, and Pair-GRAPE-S when applicable had the highest percent utilities, producing optimal solutions for all trials. SDA's solutions were also optimal for most trials; however, SDA's percent utilities for 100 robot collectives with ten service types, five services per robot, and 50\% tasks were slightly lower (i.e., 99.99\% median, min. $=$ 99.87\%, max. $=$ 100.0\%).

Overall, each of the algorithms produced high quality solutions. GRAPE-S's percent utilities were lower than RACHNA$_{dt}$'s in the worst case, but equivalent for most independent variable values. Additionally, GRAPE-S's percent utilities were worse than SDA's in the worst case, but equivalent or better on average. Nevertheless, the algorithms' utilities differed significantly for collectives with ten service types and five services per robot ($\chi_r^2$ ($n=200$) $=$ 39.00, $p<0.01$). Specifically, GRAPE-S's utilities were significantly worse than the other algorithms ($p<0.01$), although largely near-optimal. No other significant differences were detected. The algorithms' percent utilities also differed significantly overall ($\chi_r^2$ ($n=800$) $=$ 17.63, $p<0.01$). GRAPE-S's utilities were significantly worse than SDA's ($p<0.01$), as well as RACHNA$_{dt}$'s ($p<0.01$), but no significant differences were identified between SDA and RACHNA$_{dt}$.

\subsection{Discussion}

The centralized evaluation provided an initial assessment of whether GRAPE-S and Pair-GRAPE-S are viable for very large collectives in highly dynamic domains, where a viable algorithm produces high utility solutions in near real-time while using minimal communication. GRAPE-S and Pair-GRAPE-S were also compared to the auction baselines SDA and RACHNA$_{dt}$.


Recall that Pair-GRAPE-S is equivalent to GRAPE-S for homogeneous and single-service collectives (i.e., all collective compositions except ten service types, five services per robot). Pair-GRAPE-S best met the near real-time target ($H_1$), being the only algorithm with $<5$ min median runtimes for all collective sizes, compositions, and percent tasks. Pair-GRAPE-S was closely followed by GRAPE-S. Both algorithms did have rare outliers that exceeded the target, but they were not exceptionally unreasonable for near real-time domains; however, the maximum runtimes for SDA and RACHNA$_{dt}$ were far too long. Notably, SDA and RACHNA$_{dt}$'s runtimes were especially long for collectives with five services per robot. The collective composition with ten service types and five services per robot is the most representative of real-world heterogeneous robot collectives that must provide multiple services simultaneously, making SDA's and RACHNA$_{dt}$'s runtimes less viable for near real-time domains than GRAPE-S's and Pair-GRAPE-S's.

GRAPE-S and Pair-GRAPE-S were also hypothesized ($H_2$) to be more suitable for communication constrained domains than the baseline algorithms, which was not supported. GRAPE-S and Pair-GRAPE-S only had low communication requirements for 100 robots. Additionally, their multiple gigabyte communication requirements with 1000 robots far exceed a deployed networks expected capacity. GRAPE-S's average communication requirement and Pair-GRAPE-S's worst case communication requirement were also especially high with ten services and five services per robot, which is the composition most representative of real-world heterogeneous collectives.  This communication metric is the main barrier to meeting the domain requirements. SDA consistently required the least communication, while RACHNA$_{dt}$ scaled poorly with increased collective size and percent tasks. Thus, SDA is the most suitable for deployed networks, not Pair-GRAPE-S or GRAPE-S.

GRAPE-S's and Pair-GRAPE-S's utilities were expected to be comparable to the baselines ($H_3$), which was largely supported. Pair-GRAPE-S performed as well as RACHNA$_{dt}$ and better than SDA. GRAPE-S's solution quality was somewhat lower, being the only algorithm to produce a solution below the $>95$\% threshold; however, it appears to be a worst case instance. The target domains' near real-time requirement may also result in a preference for somewhat lower quality solutions in near real-time, over to optimal solutions on a much longer time scale. 

The hypothesis ($H_4$) that Pair-GRAPE-S's utilities are higher than GRAPE-S's was supported. Unlike GRAPE-S, Pair-GRAPE-S produced optimal solutions in near real-time, regardless of collective composition. Given these results, Pair-GRAPE-S is likely viable for near real-time domains when sufficient communication infrastructure is available; however, deployed networks will require reducing Pair-GRAPE-S's communication complexity.

Utility was not the only aspect in which GRAPE and Pair-GRAPE-S differed notably. Pair-GRAPE-S exists to improve solution quality by running the pairwise stability step whenever GRAPE-S produces a suboptimal solution; however, Pair-GRAPE-S required the pairwise stabililty step far more often than GRAPE-S produced subotpimal (i.e., Pair-GRAPE-S's internal GRAPE-S component performed worse than GRAPE-S on its own). A likely explanation is that Pair-GRAPE-S's implementation limited the number of coalition members to the maximum number of robots required for task performance, facilitating determining if the pairwise stability step was needed. This limit can reduce GRAPE-S's ability to produce an optimal solution by preventing some coalition selections. This limitation can be avoided by removing the coalition size limit and instead removing excess robots from coalitions immediately before performing the pairwise stability step; however, Pair-GRAPE-S was able to correct for the limitation and produce optimal solutions. Additionally, the limit resulted in Pair-GRAPE-S having even faster runtimes and less communication than GRAPE-S. These results suggest that capping the coalition sizes during Pair-GRAPE-S may be useful in practice.

\section{Distributed Evaluation}

The centralized simulation results were generated with each robots' computation performed iteratively; however, real-world robots will perform their computation asynchronously and in parallel. The distributed evaluation assesses algorithm viability in more realistic conditions, using Oregon State University's high performance computing cluster (HPCC): a heterogeneous set of compute nodes connected by a Mellanox EDR InfiniBand network. Each robot was represented by a single core with two threads. The first thread ran the coalition formation algorithms, and the second received and processed communications. Communication between processes used the Message Passing Interface (MPI) per the HPCC's standard. All communication was non-blocking to model a robot using UDP broadcasts. The simulations modeled a fully-connected network topology; however, communication was not instantaneous.

The algorithms analyzed are GRAPE-S, Pair-GRAPE-S, and SDA (Table \ref{table:ind_variables_achievable2}). RACHNA$_{dt}$ was eliminated from consideration, due to its multiple hour worst case runtimes. Parallelization can potentially decrease runtime, but most of RACHNA$_{dt}$'s computation occurs on the task and service agents, which are a small portion of the collective; thus, the potential benefit of distribution is limited. RACHNA$_{dt}$'s synchronized bidding rounds also limit the potential benefits, as achieving synchronicity in an asynchronous environment typically requires communication, which can be time-consuming. 

\begin{table}[h]
\normalsize
    \centering
    \caption{Distributed evaluation: independent variables}
    \begin{tabular}{|c|l|}
        \hline
        Collective Size & 100, 500 \\
        \hline
        Percent Tasks & 1, 10, 50\\
        \hline
        Service Types: Per Robot & 5:1 , 5:5, 10:1, 10:5\\
        \hline
    \end{tabular}
    \label{table:ind_variables_achievable2}
  
\end{table}

The coalition formation problems considered are a subset of the achievable mission problems from the centralized evaluation. The collective sizes are reduced from 100, 500, and 1000 to only 100 and 500, due to the HPCC's inability to support 1000 robot collectives. This reduction decreases the number of trials to 500 per algorithm. Additionally, the maximum runtime is one hour, as this limit is too long for near real-time domains, but permits efficient use of the HPCC's resources. The remaining independent variables, dependent variables, and hypotheses are identical to the centralized evaluation.

GRAPE-S's and Pair-GRAPE-S's runtimes were also expected to benefit substantially from parallel computation, as computation is divided evenly among all robots in the collective. SDA's runtimes were expected to benefit less, as, like RACHNA$_{dt}$, most computation occurs on the task and service agents, which are a small portion of the collective. All algorithms' communication and solution quality were expected to be minimally impacted, although both may degrade slightly due to the introduction of asynchronous communication.

\subsection{Results}

GRAPE-S and SDA produced solutions for all 500 trials. Recall that Pair-GRAPE-S differs from GRAPE-S only when the pairwise stability step is needed (i.e., Pair-GRAPE-S's initial solution, produced by performing GRAPE-S, is suboptimal). The only collective composition that required the pairwise stability step was ten service types and five services per robot. Pair-GRAPE-S produced solutions for all 125 trials with this collective composition and the required pairwise stability step in 80 trials. The results are presented by dependent variable. Boxplots and non-parametric statistical analyses were used, as the results are not normally distributed. 


Each algorithms' performance with respect to the other independent variables  was assessed individually. Mann-Whitney-Wilcoxon tests compared the numbers of service types and the services per robot ($n=250$), while Kruskal-Wallis analysis with Mann-Whitney-Wilcoxon post-hoc tests compared across the collective sizes and percent tasks. The service types and services per robot analysis was only applicable to GRAPE-S and SDA, not Pair-GRAPE-S. Additionally, the collective size analysis considered only 10\%-50\% tasks, and the percent task analysis considered only 500 robot collectives, as 100 robots with 1\% tasks was not evaluated.

The Wilcoxon signed-rank test assessed if the results for each dependent variable differed significantly between GRAPE-S and SDA. This analysis considered all 500 independent variable combinations. Comparison between GRAPE-S, Pair-GRAPE-S, and SDA with ten service types and five services per robot used Friedman tests with Conover post-hoc analyses.

\subsubsection{Runtime Results}

\begin{figure*}[b]
    \subfloat[Five service types, one service per robot. \label{subfig:druntime1}]{%
      \includegraphics[width=0.48\linewidth]{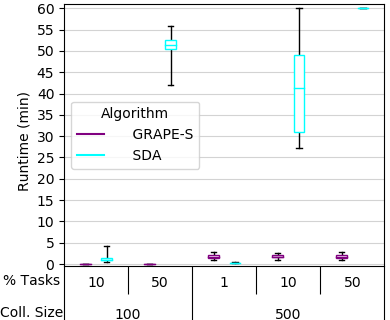}
    } 
    \hfill
    \subfloat[Five service types, five services per robot. \label{subfig:druntime2}]{%
      \includegraphics[width=0.49\linewidth]{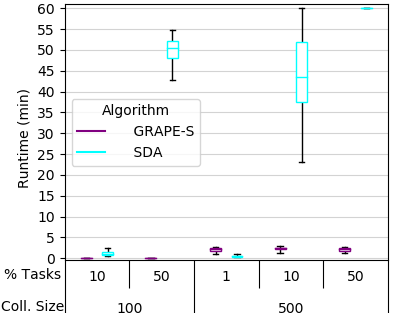}
    } 
    \hfill
     \subfloat[Ten service types, one service per robot. \label{subfig:druntime3}]{%
      \includegraphics[width=0.49\linewidth]{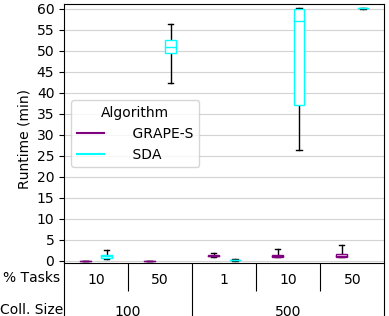}
    } 
    \hfill
    \subfloat[Ten service types, five services per robot. \label{subfig:druntime4}]{%
      \includegraphics[width=0.49\linewidth]{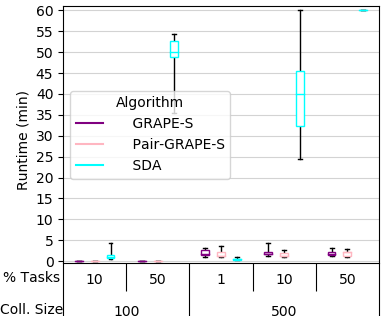}
    } 
    \caption{Distributed algorithm runtimes by collective size, percent tasks, number of services, and number of services per robot. Pair-GRAPE-S is identical to GRAPE-S in Figures \ref{subfig:runtime1}-\ref{subfig:runtime3} and is not shown separately }
    \label{fig:druntimes_achievable}
\end{figure*}

GRAPE-S's runtimes were well within the 5 min target for all 500 trials (Fig, \ref{fig:druntimes_achievable}). Only one trial had a  $>4$ min runtime (i.e., 4 min 22 s with ten service, five services per robot, 500 robots, and 50\% tasks). Additionally, only three trials had 3-4 min runtimes, with the other 496 trials completing in $<3$ min. Although GRAPE-S's runtimes were consistently low, they did increase significantly with the collective size ($p<0.01$), as well as the number of service types ($p<0.01)$; however, no other significant differences were detected.

Recall that Pair-GRAPE-S is identical to GRAPE-S for homogeneous collectives and collectives with one service per robot (Figures \ref{subfig:druntime1}-\ref{subfig:druntime3}). For collectives with ten service types and five services per robot (Figure \ref{subfig:druntime4}), Pair-GRAPE-S performed similarly to GRAPE-S, completing all trials well within the target. The longest runtime was only 3 min 33 s, which occurred with 500 robots, 1\% tasks, five service types, and one service per robot. All other trials completed in $<3$ min. Like GRAPE-S, Pair-GRAPE-S's runtimes did increase significantly with the collective size ($p<0.01$), and no difference was found across the percent tasks.

SDA's runtimes with 500 robots and 1\% tasks were slightly faster than GRAP-E-S's and Pair-GRAPE-S's. Additionally, SDA's runtimes with 100 robots and 10\% tasks were not meaningfully longer, with all trials completing in $<5$ min; however, SDA's runtimes increased substantially with the collective size and percent tasks. 100 robot collectives with 50\% tasks required 35-57 min, while 500 robots with 10\% tasks required $\geq25$ min and sometimes exceeded the one hour limit. No trial with 500 robots and 50\% tasks completed within the time limit. This trend is consistent with SDA's  $O(mn^4u_{max}/\epsilon_{dec}$) computational complexity.

SDA's results were largely consistent across the different collective compositions with the exception of the 500 robot, 10\% task trials. The median runtimes with 500 robots and 10\% tasks differed by $<5$ min for collectives with five service types with one service per robot, five service types with five services per robot, and ten services with five services per robot. Runtimes for five services and one service per robot were somewhat more variable; however, runtimes with ten service types and five services per robot were much longer. A possible explanation is that ten service types corresponds to the highest number of service agents, which increases the time required for coordination. Additionally, this collective composition has the lowest redundancy in robot capabilities, which may increase the number of bidding rounds if a task has to wait for a robot with a specific capability to have its salary lowered. It is also worth noting that 500 robot collectives with 50\% tasks may exhibit similar trends in the absence of the one hour runtime limit.

Overall, SDA's runtimes increased significantly with the collective size ($p<0.01$) and differed significantly with the number of services per robot ($p=0.04$). SDA's runtimes also increased significantly with the percent tasks ($H$ ($n=25$) $=$ 99.87, $p<0.01$). Specifically, significant differences were found between 1\% tasks and 10\% tasks ($p<0.01$), as well as 1\% and 50\% tasks ($p<0.01$). No significant difference between 10\% and 50\% tasks existed.

GRAPE-S's and SDA's runtimes differed significantly ($p<0.01$), considering all collective compositions. The difference between GRAPE-S, Pair-GRAPE-S, and SDA with ten services and five services per robot was not mathematically significant; however, it was substantially different for 100 robot collectives with 50\% tasks, as well as 500 robot collectives with 10\%-50\% tasks.


\subsubsection{Communication Results}

\begin{figure*}[b]
    \subfloat[Five service types, one service per robot. \label{subfig:dcomms1}]{%
      \includegraphics[width=0.49\linewidth]{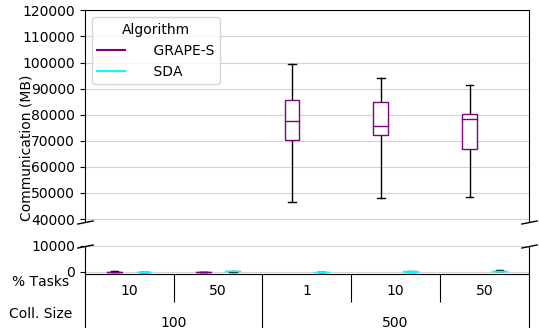}
    } 
    \hfill
    \subfloat[Five service types, five services per robot. \label{subfig:dcomms2}]{%
      \includegraphics[width=0.49\linewidth]{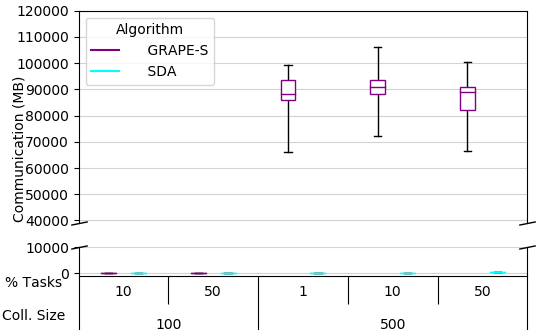}
    } 
    \hfill
     \subfloat[Ten service types, one service per robot. \label{subfig:dcomms3}]{%
      \includegraphics[width=0.49\linewidth]{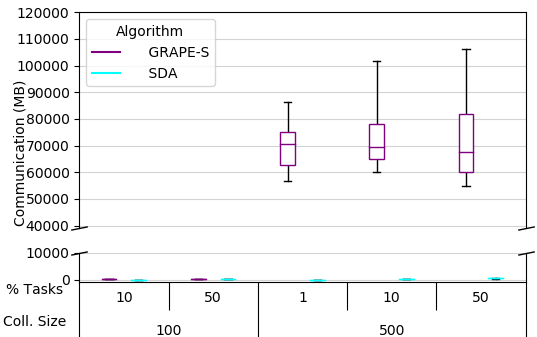}
    } 
    \hfill
    \subfloat[Ten service types, five services per robot. \label{subfig:dcomms4}]{%
      \includegraphics[width=0.49\linewidth]{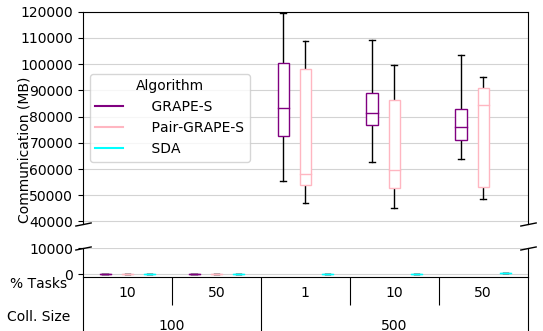}
    } 
    \caption{Distributed algorithm's total communication by collective size, percent tasks, number of services, and number of services per robot. Pair-GRAPE-S is identical to GRAPE-S in Figures \ref{subfig:comms1}-\ref{subfig:comms3} and is not shown separately }
    \label{fig:dcomms_achievable}
\end{figure*}

GRAPE-S's communication requirement was reasonable for 100 robot collectives (i.e., $<100$ MB in total), as shown in Fig. \ref{fig:dcomms_achievable}; however, GRAPE-S required as much as 50-120 GB of communication for for 500 robot collectives, which is very high for a deployed network. The highest and most variable communication requirement occurred with ten services and five services per robot, consistent with the centralized results. The communication increase with collective size was significant ($p<0.01$), as was the increase with the number of services per agent ($p<0.01$). No other significant differences existed.


Pair-GRAPE-S was equivalent to GRAPE-S for all collective compositions except ten service types and five services per robot. Pair-GRAPE-S's communication for this collective composition was also similar to GRAPE-S (i.e., $\leq 110$ MB for 100 robot collectives, 45-110 GB with 500 robots). The increase with the collective size was significant ($p<0.01$), but no difference across the percent tasks was found.

SDA's communication requirement was consistently low, unlike GRAPE-S and Pair-GRAPE-S. All SDA trials required $<500$ MB. Nevertheless, the communication increased significantly from 100 to 500 robots ($p<0.1$), as well as with the percent tasks ($H$ ($n=25$) $=$ 231.92, $p<0.01$). All post hoc analyses were significant.

Overall, SDA's communication was significantly lower than SDA's ($p<0.01$). The algorithms' communication also differed significantly for collectives with ten service type and five services per robots  ($H$ ($n=25$) $=$ 116.42, $p<0.01$). Specifically, SDA's communication was significantly lower than GRAPE-S's ($p<0.01$) and Pair-GRAPE-S's ($p<0.01$). No significant difference between GRAPE-S and Pair-GRAPE-S was identified.

\subsubsection{Utility Results}

GRAPE-S's produced optimal 100\% utility solutions for 495 of 500 trials. The five subpotimal trials all incorporated collectives with ten service types and five services per robot (Table \ref{table:dutilities_achievable_grapes}). Although the solutions for these trials were suboptimal, their utilities were still high. The lowest solution utility, 96.7\%, was the only utility $<98$\%; thus, all of GRAPE-S's solutions were at least near-optimal. Nevertheless, the presence of the suboptimal trials meant that GRAPE-S's utilities decreased significantly with increased service types ($p<0.01$) and services per agent ($p<0.01$). GRAPE-S's utilities also differed significantly across the percent tasks ($H$ ($n=25$) $=$ 364.20, $p<0.01$) with significant post hoc analyses ($p<0.01$). No significant differences were detected between 100 and 500 robots. Unlike GRAPE-S, Pair-GRAPE-S produced optimal solutions for all trials.


\begin{table}[h]
    \centering
    \normalsize
        \caption{GRAPE-S's percent utility statistics with ten service types and five services per robot. 1\% tasks had 100\% utility, as did all other collective compositions}
    \begin{tabular}{|c|c|c|}
    \hline
          Percent & Collective & Median \\
         Tasks & Size & (Minimum, Maximum)\\
         \hline
         \multirow{2}{*}{10} & 100 & 100.0 (100, 100.0)\\
         & 500 & 100.0 (96.7, 100.0) \\
         \hline
          \multirow{2}{*}{50}&  100 & 100.0 (98.8, 100.0)\\
         & 500 & 100.0 (99.3, 100.0)\\
         \hline
    \end{tabular}
    \label{table:dutilities_achievable_grapes}
\end{table}

SDA's percent utilities were optimal or near-optimal for the majority of of the 100 robot trials, as well as trials incorporating 500 robot collectives and 1\%-10\% tasks (Table \ref{tab:sda_distributed_utilities}); however, SDA's worst case utilities for these independent variable combinations were unacceptably low. The low worst-case utilities with 500 robots and 10\% tasks may be attributed to the fact that some trials reached the time limit without completing, resulting in fewer tasks being assigned coalitions; however, all other low worst-case utilities occurred for trials that completed. Lost or delayed messages likely contributed to the poor results, as they can prevent bids from occurring or assign the same robot to multiple tasks. SDA's utilities with 500 robots and 50\% tasks were generally even lower (i.e., $<5$\%), due to SDA's failure to complete within the time limit. SDA's percent utilities decreased significantly with the collective size ($p<0.01$), and differed significantly with the number of services per robot ($p<0.01$). The utilities also decreased significantly with the percent tasks ($H$ ($n=25$) $=$ 16.12, $p<0.01$). Specifically, there were significant differences between 1\% and 10\% tasks ($p<0.01)$, as well as 1\% and 50\% tasks ($p<0.01$). No other significant differences were detected.


\begin{table}[h]
    \centering
    \normalsize
    \caption{SDA's percent utilities. Percent utilities $<95$\% are noted in \textcolor{red}{red}}
    \label{tab:sda_distributed_utilities}
    \begin{NiceTabular}{|c|c|c|c||c|}
        \hline
       Service & Services & Percent & Collective & Median \\
        Types & Per Robot & Tasks & Size &  (Minimum, Maximum)\\
        \hline
        \hline
        \multirow{5}{*}{5} & \multirow{5}{*}{1} & 1 & 500 & 100 (\textcolor{red}{73.7}, 100) \\\cline{3-5}
        &&\multirow{2}{*}{10} & 100  & 100 (\textcolor{red}{86.4}, 100)\\
        & & & 500 & 100 (\textcolor{red}{58.5}, 100)\\\cline{3-5}
        & & \multirow{2}{*}{50} & 100 & 100 (97.1, 100)\\
        & & & 500  & \textcolor{red}{3.9 (1.6, 8.2)} \\
        \hline
        \multirow{5}{*}{5} & \multirow{5}{*}{5} & 1 & 500 & 100 (100, 100) \\\cline{3-5}
        &&\multirow{2}{*}{10} & 100  &  100 (\textcolor{red}{94.1}, 100)\\
        & & & 500 & 100 (\textcolor{red}{6.8}, 100) \\\cline{3-5}
        & & \multirow{2}{*}{50} & 100 & 100 (\textcolor{red}{93.5}, 100)\\
        & & & 500  & \textcolor{red}{3.8 (0.8, 6.4)}\\
        \hline
        \multirow{5}{*}{10} & \multirow{5}{*}{1} & 1 & 500 & 100 (100, 100) \\\cline{3-5}
        &&\multirow{2}{*}{10} & 100  & 100 (\textcolor{red}{90.2}, 100)\\
        & & & 500 & 99.8 (\textcolor{red}{14.2}, 100) \\\cline{3-5}
        & & \multirow{2}{*}{50} & 100 &  100 (100, 100)\\
        & & & 500  & \textcolor{red}{3.2 (1.5, 6.4)}\\
        \hline
        \multirow{5}{*}{10} & \multirow{5}{*}{5} & 1 & 500 & 100 (\textcolor{red}{64.2}, 100) \\\cline{3-5}
        &&\multirow{2}{*}{10} & 100  & 100 (\textcolor{red}{63.4}, 100) \\
        & & & 500 & 100 (\textcolor{red}{3.0}, 100) \\\cline{3-5}
        & & \multirow{2}{*}{50} & 100 &  99.8 (96.5, 100)\\
        & & & 500  & \textcolor{red}{4.5 (2.4, 27.4)}\\
        \hline
       
    \end{NiceTabular}
\end{table}

Overall, SDA's utilities were significantly lower than GRAPE-S's ($p<0.01$). No significant difference existed when only collectives ten service types and five services per robot were considered.

\subsection{Discussion}

The distributed evaluation assessed GRAPE-S's and Pair-GRAPE-S's ability to produce near real-time collective coalition formation solutions in a more realistic asynchronous setting. Minizing communication was also preferred.

GRAPE-S and Pair-GRAPE-S had better runtimes than SDA, as expected ($H_1$). Pair-GRAPE-S's runtimes were the fastest and least variable; however, there was little practical difference between GRAPE-S and Pair-GRAPE-S, as both consistently met the 5 min runtime target. Either algorithm produced solutions sufficiently quickly for near real-time domains.

The hypothesis that GRAPE-S and Pair-GRAPE-S require less communication than SDA was not supported ($H_2$). SDA's communication requirement was relatively low, while GRAPE-S's and Pair-GRAPE-S's were excessive. There was little practical difference between  Pair-GRAPE-S and GRAPE-S.

GRAPE-S's and Pair-GRAPE-S's utilities were also expected to be comparable to SDA's ($H_3$), but they were actually much better in many cases. SDA's single digit worst-case utilities are much too low to support practical real-world deployments. Pair-GRAPE-S's utilities are preferable to GRAPE-S's ($H_4$); however, both GRAPE-S and Pair-GRAPE-S produced sufficiently high quality solutions for real-world applications.

Overall, Pair-GRAPE-S's performance was the best suited for collective coalition formation,  given sufficient communication infrastructure. GRAPE-S is also viable, but at the cost of lower utility solutions with no benefit to other relevant metrics. SDA, despite its low communication cost, is not well suited to near real-time domains, due to its long runtimes and poor solution quality.

\section{Discussion}

Near real-time domains, such as disaster response, require low communication algorithms that produce high quality task allocations (i.e., $95$\% utility) quickly (i.e., $\approx 5$ min). GRAPE-S leverages GRAPE's fast runtimes with large collectives, but also permits multiple services. Additionally, Pair-GRAPE-S was introduced to improve solution quality for heterogeneous, multiple service collectives. GRAPE-S's and Pair-GRAPE-S's suitablity across the domain metrics were compared to the baseline auction protocols, SDA and RACHNA$_{dt}$ in centralized simulations. GRAPE-S, Pair-GRAPE-S, and SDA were also evaluated in distributed simulations.

RACHNA$_{dt}$'s strength is its high percent utilities in a centralized setting. These results are well suited to pre-mission planning, where multiple hour runtimes are permissible. However, RACHNA$_{dt}$'s runtimes scaled poorly with the number of services per robot (i.e., the number of robots per service agent) and were too long for near real-time domains. These runtimes cannot be sufficiently improved through distribution, as the service agents are a small portion of the collective. RACHNA$_{dt}$ is eliminated from further consideration, as multiple service robots are needed to enable a wider range of collective applications.

SDA performed relatively well in a centralized setting, despite somewhat long worst-case runtimes; however, distributed SDA's performance was much worse, with lower utilities and longer runtimes. These results reflect two key algorithm limitations. The first is that SDA requires substantial time to coordinate bids and communicate robot pricing when communication is not instantaneous. It is possible that increasing the bid decrement can reduce runtimes, at the cost of lower solution quality; however, solution quality was already sometimes too low, even when the runtime limit was not reached. The other limitation is that SDA is not robust to lost or delayed messages, as reflected in the distributed implementation's lower solution quality. This limitation is not specific to SDA, but fundamental to auctions. Buyers and sellers must have up-to-date information about the price of goods, which is hindered by asynchronicity, non-instantaneous communication, and message loss. Collectives' scale can exacerbate all of these issues; thus, auctions are likely best reserved for smaller multiple robot systems or centralized pre-mission planning. Other approaches are needed for collective coalition formation. 

GRAPE-S and Pair-GRAPE-S were the best suited to near real-time collective coalition formation, satisfying the runtime and utility requirements in both the centralized and distributed simulations. Both GRAPE-S variants' runtimes improved when distributed meaning that the advantage of parallelizing computation outweighed the time cost for non-instantaneous communication. This result reflects the fact that GRAPE-S's and Pair-GRAPE-S's computation is distributed evenly across the entire collective. Both algorithms' centralized and distributed utilities were also similar, demonstrating that they are at least somewhat resilient to delayed or dropped messages. Further study is needed to determine the limits of this resilience.

Communication was GRAPE-S and Pair-GRAPE-S's main limitation. Both algorithms had high communication requirements for distributed 500 robot collectives, likely exceeding what a deployed network can provide. Notably, the distributed implementations also required far more communication than the centralized implementations, due to distributed iterations being asynchronous. A robot in a distributed setting may not receive all relevant messages in the same iteration and waste the next iteration choosing a coalition based on outdated information. Acting on outdated information is likely unavoidable in highly distributed, asynchronous applications; however, communication reduction may be possible by allowing for multiple modifying agents or using heuristics to determine when communication is most beneficial. Additionally, the algorithms' current communication requirements can likely be supported by permanent communication infrastructure (e.g., cellular), which can sometimes be available.

Overall, the results demonstrate that hedonic games can produce high utility coalition formation solutions in near real-time for very large collectives, given sufficient communication resources, while prior approaches were not viable for near real-time domains. Additionally, GRAPE-S and Pair-GRAPE-S make near real-time collective coalition possible for multiple service collectives, which is necessary to provide coalition formation capabilities for sophisticated tasks and missions.

\section{Conclusion}

Robotic collectives can provide disaster response operations with the ability to perform numerous tasks simultaneously across a large spatial area. Coalition formation is needed to assign robots to effective task teams. A coalition formation algorithm must produce near-optimal solutions in near real-time, so that the collective can respond quickly to tasks as they arise. Additionally, communication must be minimized and distributed. This manuscript introduced the GRAPE-S and Pair-GRAPE-S algorithms, which integrated GRAPE with the services model in order to achieve GRAPE's fast runtimes for collectives with multiple service types. Both algorithms successfully produced optimal or near-optimal solutions within the 5 min target for distributed collectives, given sufficient communication resources, satisfying the utility and runtime requirements for near real-time domains. Additionally, these algorithms transferred well from centralized to distributed systems, unlike the auction baselines, making GRAPE-S and Pair-GRAPE-S the first algorithms demonstrated to support near real-time coalition formation for very large, distributed collectives with multiple services. Supporting such coalition formation can enable very large collectives to leverage robots' diverse capabilities effectively, enabling complex missions and task performance.

\backmatter

\bmhead{Acknowledgments}

Diehl's Ph.D. was supported by an ARCS Foundation Scholar award, the Office of Naval Research, and the Defense Advanced Research Projects Agency. The views, opinions, and findings expressed are those of the author and are not to be interpreted as representing the official views or policies of the Department of Defense or the U.S. Government.

\section*{Declarations}

\subsection*{Competing Interests and Funding}

Diehl has received research support from the Defense Advanced Research Projects Agency and the Office of Naval Research. The authors have no relevant financial or non-financial interests to disclose.

\subsection*{Data Availability}

The datasets generated during the current study are available from the corresponding author on reasonable request.

\subsection*{Ethics Approval}

Not applicable.

\subsection*{Consent to participate}

Not applicable.

\subsection*{Consent for publication}

 Not applicable.

 \subsection*{Author's contributions}

Both authors contributed to the study conception and design. Coding, data collection, and analysis were performed by G.D., supervised by J.A.A.. The first draft of the manuscript was written by G.D., and J.A.A. provided feedback on previous versions of the manuscript. Both authors read and approved the final manuscript.


\bibliography{references}


\end{document}